\documentclass[preprint, showpacs,preprintnumbers,amsmath,amssymb,showkeys]{revtex4}
\usepackage{amsmath}
\usepackage{bm}
\usepackage{graphicx}

\newcommand{\be}{\begin{equation}}
\newcommand{\ee}{\end{equation}}
\newcommand{\bea}{\begin{eqnarray}}
\newcommand{\eea}{\end{eqnarray}}
\newcommand{\bc}{\begin{center}}
\newcommand{\ec}{\end{center}}

\begin{document}

\title{Skyrmions, Hadrons and isospin chemical potential}
\author{M. Loewe}
\email{mloewe@fis.puc.cl} \affiliation{Facultad de F\'\i sica,
Pontificia Universidad Cat\'olica de Chile,\\ Casilla 306,
Santiago 22, Chile.}
\author{S. Mendizabal}
\email{smendiza@fis.puc.cl} \affiliation{Facultad de F\'\i sica,
Pontificia Universidad Cat\'olica de Chile,\\ Casilla 306,
Santiago 22, Chile.}
\author{J.C. Rojas}
\email{jurojas@ucn.cl} \affiliation{Departamento de F\'{\i}sica,
Universidad Cat\'{o}lica del Norte, Casilla 1280, Antofagasta,
Chile}

\begin{abstract}
Using the Hamiltonian formulation, in terms of collective variables,
we explore the evolution of different skyrmionic parameters as
function of the isospin chemical potential ($\mu$), such as the
energy density, the charge density, the isoscalar radius and the
isoscalar magnetic radius. We found that the radii start to grow
very fast for $\mu \gtrsim 140$ MeV, suggesting the occurrence of a
phase transition.
\end{abstract}

\maketitle

The skyrmionic approach to baryon dynamics is an interesting attempt
to discuss the occurrence of phase transitions induced by
temperature and/or density effects in the hadronic sector
\cite{lmr3}. The idea is to study the stability properties of such
objects by an analysis of their mass behavior and the spatial
extension, i.e. mean square radius, associated to different
currents. Recently this idea has been extrapolated also to the
analysis of Skyrmions in curved spaces \cite{ati-suth}.

In a previous article, we discussed the static properties of the
Skyrmion solutions in the presence of a finite isospin chemical
potential $\mu$ \cite{lmr3}. We were able, among other results, to
find a critical chemical potential $\mu_c=222.8$ MeV, where the
Skyrmion mass vanishes.

In this article we will extend our analysis by using the Hamiltonian
formulation introduced by Adkins et al. \cite{anw}. Note that the
identification of baryons as Skyrmion states emerges only in the
Hamiltonian formalism. The difference between the static analysis
and the present one, is that now we are able to look into the energy
spectrum of non strange nucleons as function of $\mu$. In the static
approach we cannot distinguish between different nucleon states.

Using collective coordinates $a_i$ $(i=1..4)$, following \cite{anw},
we establish the Lagrangian in terms of $a_i$ and $\dot{a}_i$. Then
we infer the Hamiltonian operator which gives us the energy levels
of the nucleon, the charge densities, the isoscalar mean square
radius and the magnetic mean square radius. In particular, we
emphasize the growing behavior of the mean squared radius,
associated to different conserved currents, as function of $\mu$,
suggesting the occurrence of a phase transition.

The Skyrme lagrangian with isospin chemical potential ($\mu$) is
given by

\bea \label{lagra}{\cal L} &=& \frac{F_{\pi}^2}{16} Tr\left[
D_{\mu}U D^{\mu}U^{\dagger} \right] \nonumber \\ &&+
\frac{1}{32e^2}Tr \left[ (D_{\mu}U)U^{\dagger},
(D_{\nu}U)U^{\dagger} \right]^2. \label{Lag}\eea

As usual, the isospin chemical potential is introduced through the
covariant derivative \cite{weldon,actor}

\be D_{\mu}=\partial_{\mu}-i\frac{\mu}{2}
\left[\sigma_3,U\right]\delta_{\mu,0}. \ee

In ref \cite{lmr3}, the static solution $U_0(r)$ of  (\ref{lagra})
was found using a radial symmetric ``Hedgehog" ansatz \cite{anw},
finding the minimum of the Skyrmion mass with the parametrization

\be U_0=\exp \left[i F(r) \sigma_i \hat{n}_i \right], \ee

\noindent where the $\sigma_i$'s are the Pauli matrices, $\hat{n}_i$
denotes the spatial unitary vector and F(r) is a numerically
determined function, that satisfies $F(0)=\pi$ and $F(\infty)=0$.
When the isospin chemical potential is turned on, the profile
acquires also a dependence on $\mu$.

\begin{equation}
F(r)\rightarrow F(r,\mu).
\end{equation}

With this profile, the energy density acquires the following form
\cite{lmr3}

\be M_{\mu}=M_{\mu=0}- \frac{\mu^2}{4 e^3 F_{\pi}} I_2 -
\frac{\mu^2}{32 e^3 F_{\pi}} I_4, \label{mmu}\ee

\noindent with

\bea M_{\mu=0} &=& \frac{F_{\pi}}{4 e} \left\{ 4 \pi
\int_0^{\infty}d\hat{r} \left[ \frac{\hat{r}^2}{2} \left(
\frac{dF}{d\hat{r}}\right)^2 + \sin^2F\right]\right. \nonumber
\\  && + 4 \pi \int_0^{\infty}d\hat{r}
\frac{\sin^2F}{\hat{r}^2} \times \nonumber \\ & &  \left. \left[
4\hat{r}^2 \left( \frac{dF}{d\hat{r}}\right)^2+2\sin^2F\right]
\right\},\eea

\noindent where we introduced the dimensionless parameter
\newline $\hat{r}=e F_{\pi} r$ and the integrals $I_2$, $I_4$

\bea I_2 &=& \int d^3\hat{r} \mathrm{Tr} \left[ \sigma_0
-U\sigma_3U^{\dagger}\sigma_3\right],
\nonumber \\
I_4 &=&  \int d^3\hat{r} \mathrm{Tr} \left[ \varrho,L_{\nu}
\right]^2, \label{Ies}\eea

\noindent with $\varrho \equiv \sigma_3-U\sigma_3 U^{\dagger}$,
$\sigma_0$ is the $2\times 2$ identity matrix and $L_{\nu}\equiv
(\partial_{\nu}U)U^{\dagger}$. The variational equation for the
static Skyrme Lagrangian given in (\ref{Lag}), allows us to find a
numerical solution for the profile $F(r)$ \cite{lmr3}. As it was
already mentioned, the mass of the Skyrmion vanishes for a critical
chemical potential.

As usual, in order to obtain the hadronic spectra, it is convenient
to introduce $SU(2)$ collective coordinates $A(t)$ \cite{anw}, such
that


\be U=A(t) U_0 A^{\dag}(t).\label{A}\ee

The $SU(2)$ matrix $A(t)$ is parameterized by the Pauli
$\vec{\sigma}$ matrices and the identity $\sigma_0$, according to

\be A(t)=a_0(t) \sigma_0+i\vec{a}(t) \cdot \vec{\sigma},
\label{A2}\ee

\noindent where the $a$'s obey the constraint

\be a_0^2(t)+\vec{a}^2(t)=1.\ee

Introducing (\ref{A}) and (\ref{A2}) into (\ref{Lag}), a direct (but
rather involved) computation leads us to the Lagrangian

\bea L &=& -M_{\mu}+2\lambda \left[\left(\dot{a}_0+\frac{\mu
a_3}{2}\right)^2 + \left(\dot{a}_1-\frac{\mu a_2}{2}\right)^2
\right. \nonumber \\ && + \left. \left(\dot{a}_2+\frac{\mu
a_1}{2}\right)^2 + \left(\dot{a}_3-\frac{\mu a_0}{2}\right)^2
 \right] \nonumber \\ &\equiv& -M_{\mu} + 2\lambda \left(\dot{a}_i+\mu\frac{\tilde{A}_i}{2} \right)^2, \label{lagrangian}\eea

\noindent where $\lambda=\left(2 \pi/3e^3 F_{\pi} \right) \Lambda$,
with

\be \Lambda= \int r^2 \sin^2F\left[1+4 \left(
F'^2+\frac{\sin^2F}{r^2} \right) \right]. \label{Lambda}\ee

\noindent In equation (\ref{lagrangian}) we have defined

\be \tilde{A}_i =  g_{ij}a_j,\ee

\noindent where

\be g_{ij}= \left[\begin{array}{cccc}
  0 & -1 & 0 & 0 \\
  1 & 0 & 0 & 0 \\
  0 & 0 & 0 & -1 \\
  0 & 0 & 1 & 0
\end{array}\right].\label{gij}
\ee

\noindent Explicitly,

\bea \tilde{A}_0 &=&  a_3, \;\;\;\tilde{A}_1= -a_2,\nonumber \\
\tilde{A}_2 &=& a_1, \;\;\;\tilde{A}_3=-a_0.\eea

Notice that the $\lambda$ has the same functional dependance on $F$
as the equivalent parameter defined in \cite{anw}. In our case $F$
depends also on $\mu$. $M_{\mu}$ is the chemical potential dependent
mass given in (\ref{mmu}).

From the lagrangian (\ref{lagrangian}), we get the Hamiltonian

\be H=M_{\mu}-2\lambda \mu^2+\frac{\pi^2_i}{8\lambda},  \ee

\noindent where the canonical momentum is given through a minimal
coupling

\be \pi_i=p_i-4\lambda\mu \tilde{A}_i.\ee

The Hamiltonian can be expressed in the following way

\be H = M_{\mu}+\frac{p_i^2}{8\lambda}-\mu \tilde{A}_ip_i, \ee

\noindent and considering the canonical quantization procedure $p_i
\rightarrow \hat{p}_i=-i\delta/\delta a_i$, we get

\bea H&=&M_{\mu}+\frac{1}{8\lambda}\left(-\frac{\delta^2}{\delta a_i^2}\right)+i\mu g_{ij}a_j\frac{\delta}{\delta a_i},\nonumber \\
&=&M_{\mu}-\frac{1}{8\lambda}\frac{\delta^2}{\delta a_i^2}-2\mu
\hat{I}_3, \label{hamiltonian}\eea

\noindent where $\hat{I}_3$ is the third component of the isospin
operator \cite{anw}

\be \hat{I}_k= \frac{i}{2}\left(  a_0 \frac{\delta}{\delta a_k} -
a_k \frac{\delta}{\delta a_0} -
\varepsilon_{klm}a_l\frac{\delta}{\delta a_m} \right). \ee

Following the usual procedure, we may associate a wave function to
the Skyrme Hamiltonian. In order to identify baryons in this model,
these wave functions have to be odd, i.e. $\psi(A)=-\psi(-A)$. In
particular, nucleons correspond to linear terms in the $a$'s,
whereas the quartet of $\Delta$'s are given by cubic terms.

\begin{figure}
\includegraphics[angle=0,width=0.5\textwidth]{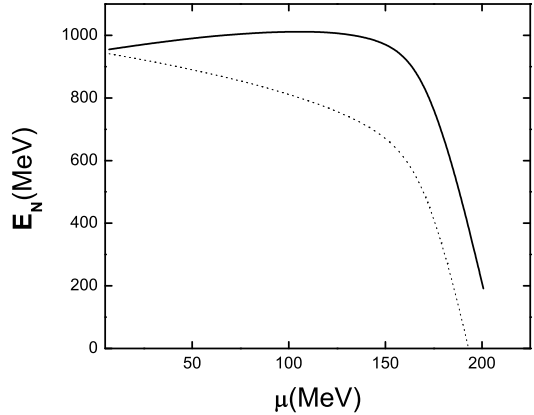}
\caption{\label{mu1mu2} The Nucleon static energy as function of the
isospin chemical potential $\mu$. The splitting between
neutron(solid line) and proton(dotted line) is due to the last term
in (\ref{hamiltonian}).}
\end{figure}

The energy spectra of nucleons as function of $\mu$ is shown in
figure \ref{mu1mu2}. We can see that an energy splitting between
neutrons and protons is induced. This Hamiltonian remind us the
Zeeman effect, where the generation of the energy spectra is broken,
by an external magnetic field. In our case the isospin chemical
potential plays the same role.

Here we are interested to establish the relation between the
relevant physical parameters and the isospin chemical potential, for
that purpose, we will consider the Baryonic, vector and axial
currents, exploring the behavior of the effective radii associated
to them.

The Skyrmion model allows the existence of different conserved
currents and their respective charges \cite{alvarez}. Using
different charge densities we may define several effective radii.
The evolution of those radii as function of chemical potential
provides information about the critical behavior close to the phase
transition, where the Skyrmion is no longer stable.

Let us first start with the topological baryonic current

\be B^{\mu}=\frac{\varepsilon^{\mu\nu\alpha\beta}}{24 \pi^2} Tr
\left[(U^{\dag}\partial_{\nu}U)(U^{\dag}\partial_{\alpha}U)(U^{\dag}\partial_{\beta}U)
\right]. \label{barionic}\ee

\noindent The baryonic charge density for the Skyrmion is given by

\be \rho_B=4 \pi r^2 B^0(r)=-\frac{2}{\pi}\sin^2F(r) F'(r). \ee

\noindent Obviously, $\int_0^{\infty}dr \rho_B=1$, independently of
the shape of the skyrmionic profile.

The isoscalar mean square radius is defined by

\be \langle r^2 \rangle_{I=0}=\int_0^{\infty}dr r^2 \rho_B.\ee

\begin{figure}
\includegraphics[angle=0,width=0.5\textwidth]{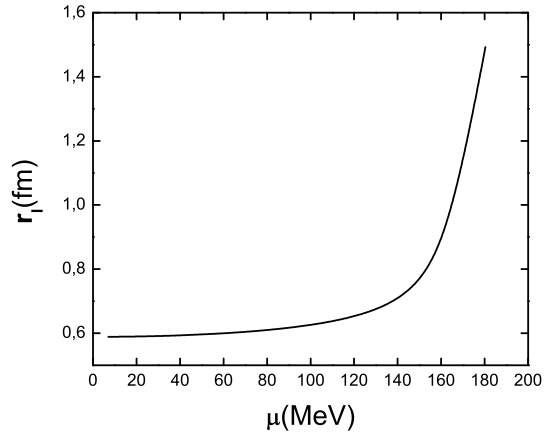}
\caption{\label{risoscalar} The isoscalar mean square radius as
function of $\mu$.}
\end{figure}

This radius seems to be quite stable up to the value of $\mu \approx
120$ MeV, starting then to grow dramatically. Although we do not
have a formal proof  that this radius diverges at a certain critical
$\mu=\mu_c$, the numerical evidence supports such claim, as it is
shown in figure \ref{risoscalar}. Divergent behavior for several
radii, associated to different currents, has also been observed in
different hadronic effective couplings as function of temperature in
the frame of thermal QCD sum rules \cite{loewe-dominguez}.

\begin{figure}
\includegraphics[angle=0,width=0.5\textwidth]{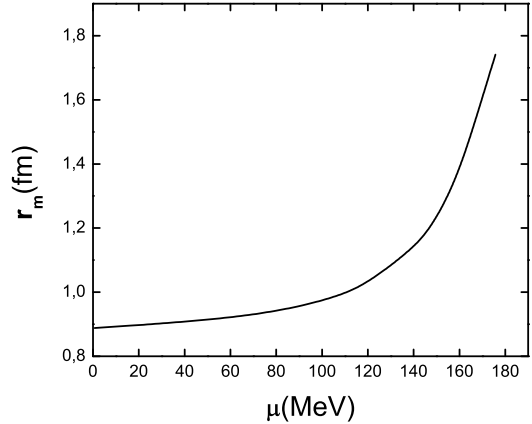}
\caption{\label{rmagnetico} The isoscalar magnetic mean square
radius as function of $\mu$. This figure suggests the same divergent
behavior as the isoscalar mean square radius with the same critical
chemical potential (figure \ref{risoscalar}).}
\end{figure}

The same kind of behavior is found for the mean square radius that
emerges from the isoscalar magnetic density

\be \rho_M^{I=0}(r)=\frac{r^2F'\sin^2F}{\int dr r^2F'\sin^2F },\ee

\noindent which is plotted in figure \ref{rmagnetico}.

This growing behavior of the mean square radius has its counterpart
in the electric charge distribution. It turns out that the proton
charge distribution becomes broader for higher chemical potentials.

Following the usual convention, we introduce the isoscalar and
isovector magnetic moments

\bea \vec{\mu}_{I=0}&=&\frac{1}{2}\int \vec{r}\times\vec{B} d^3x, \\
\vec{\mu}_{I=1}&=&\frac{1}{2}\int \vec{r}\times\vec{V}^3 d^3x,\eea

\noindent where $\vec{B}$ is the vector part of (\ref{barionic}) and
$\vec{V}$ is the Noether current associated to the vector charge.
Following \cite{anw}, we consider

\bea (\mu_{I=0})_3 &=& \frac{\langle
r^2\rangle_{I=0}}{\Lambda}\frac{e}{F_{\pi}}\frac{1}{4\pi}, \\
(\mu_{I=1})_3 &=& \frac{2}{4} \pi \frac{\Lambda}{F_{\pi}e^3}. \eea

\begin{figure}
\includegraphics[angle=0,width=0.5\textwidth]{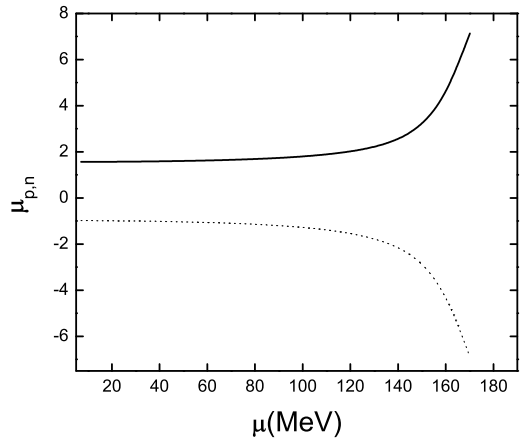}
\caption{\label{mprotneu} The magnetic moments $\mu_p$ and $\mu_n$
for the proton, (solid line) and the neutron (dotted line)
respectively, as function of $\mu$.}
\end{figure}

\begin{figure}
\includegraphics[angle=0,width=0.5\textwidth]{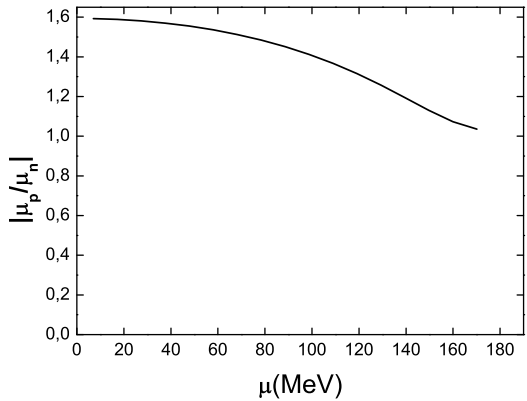}
\caption{\label{razmprotneut} The ratio $\mid \mu_p/\mu_n\mid$ as
function of $\mu$.}
\end{figure}

From the definition of the Bohr magneton for nucleons

\be \vec{\mu} = \left( \frac{g}{4M} \right) \vec{\sigma}, \ee

\noindent and making the identification $g_{I=0}=g_p+g_n$ and
$g_{I=1}=g_p-g_n$, it is possible to obtain the magnetic moments for
the proton and neutron; $\mu_p=g_p/2$ and $\mu_n=g_n/2$. The
behavior of both magnetic moments is presented in figure
\ref{mprotneu}. Besides, figure \ref{razmprotneut} shows the ratio
$\mid \mu_p/\mu_n\mid$, the figure suggests that such ratio goes to
one for $\mu \rightarrow \mu_c$.

As a conclusion, we would like to remark the divergent behavior for
the isoscalar and mean square magnetic radii as function of the
chemical potential. This suggests the occurrence of a phase
transition. In fact in the QCD sum rules approach, the radii are
phenomenological order parameters for thermal deconfinement
\cite{loewe-dominguez}. In this case, however, this behavior is
induced by the isospin chemical potential. Finally, it is
interesting to notice that the Hamiltonian approach allows us to
distinguish between the mass evolution of neutrons and protons.

\section*{ACKNOWLEDGMENTS}

The authors would like to thank financial support from
 FONDECYT grants 1051067 and 1060653.



\begin{thebibliography}{99}

\bibitem{lmr3} M. Loewe, S. Mendizabal, J.C. Rojas, Phys. Lett. B
632 (2006) 512.
\bibitem{ati-suth} Michael Atiyah, Paul Sutcliffe,
Phys. Lett. B 605 (2005) 106.
\bibitem{anw} G.S. Adkins, C.R. Nappi, E. Witten, Nucl. Phys. B
228 (1983) 552.
\bibitem{jzpr} M. Jezabek, M. Praszalowicz, {\it SKYRMIONS AND
ANOMALIES}, World Scientific, Singapore, 1978.
\bibitem{weldon} H. A. Weldon, Phys.Rev.D26 (1982) 1394.
\bibitem{actor} A. Actor,  Phys.Lett.B157 (1985) 53.
\bibitem{loewe-dominguez}  C.A. Dominguez, M. Loewe,
Phys.Lett.B481 (2000) 295;  C.A. Dominguez, M. Loewe, C. van Gend
Phys.Lett.B460 (1999) 442.
\bibitem{alvarez}  R.F. Alvarez-Estrada, F. Fernandez, J.L. Sanchez-Gomez,
V. Vento,  {\it MODELS OF HADRON STRUCTURE BASED ON QUANTUM
CHROMODYNAMICS}, Lect.Notes Phys.259 (1986), Springer Verlag.

\end{thebibliography}
\end{document}